\documentclass[10pt,conference]{IEEEtran}
\IEEEoverridecommandlockouts

\usepackage[utf8]{inputenc}
\usepackage[T1]{fontenc}
\usepackage{cite}
\usepackage{amsmath,amssymb,amsfonts}
\usepackage{algorithmic}
\usepackage{graphicx}
\usepackage{textcomp}
\usepackage{xcolor}
\usepackage{booktabs}
\usepackage{array}
\usepackage{listings}
\usepackage{url}
\usepackage{tikz}
\usetikzlibrary{arrows.meta,positioning,shapes,fit,backgrounds,calc}
\usepackage[disable]{todonotes}
\usepackage[hidelinks]{hyperref}
\usepackage{xspace}

\usepackage{flushend}   
\makeatletter
\newcommand{\linebreakand}{%
  \end{@IEEEauthorhalign}
  \hfill\mbox{}\par
  \mbox{}\hfill\begin{@IEEEauthorhalign}
}
\makeatother

\def\BibTeX{{\rm B\kern-.05em{\sc i\kern-.025em b}\kern-.08em
    T\kern-.1667em\lower.7ex\hbox{E}\kern-.125emX}}

\lstdefinelanguage{mfotl}{
  keywords={FORALL, EXISTS, ONCE, EVENTUALLY, ALWAYS, SINCE, UNTIL,
            PREV, NEXT, IMPLIES, AND, OR, NOT, TRUE, FALSE},
  keywordstyle=\bfseries,
  morecomment=[l]{//},
  morestring=[b]",
  sensitive=true
}
\lstdefinelanguage{tessla}{
  keywords={in, out, def, Events, count, last, time, filter, lift, slidingWindow, on, true, false, forall},
  keywordstyle=\bfseries,
  morecomment=[l]{//},
  sensitive=true
}
\usepackage{float}

\title{Hierarchical Security Monitoring for Edge-IoT:
       A Formal Methods Approach}

\author{%
\IEEEauthorblockN{%
\parbox{0.95\textwidth}{\centering
Nikolaos Kekatos\textsuperscript{1},
Marinelio Chintri\textsuperscript{2},
Panagiotis Katsaros\textsuperscript{3},
Alexios Lekidis\textsuperscript{4},
Tom Nianios\textsuperscript{1},
Ioannis Seitoglou\textsuperscript{2},
Stylianos Basagiannis\textsuperscript{2}}%
}
\IEEEauthorblockA{%
\parbox{0.95\textwidth}{\centering
\textsuperscript{1}Clone Systems, Cyprus,
{\footnotesize\ttfamily \{nkekatos, tnianios\}@clone-systems.com};\ %
\textsuperscript{2}International Hellenic University, Greece,
{\footnotesize\ttfamily \{basagiannis, marchint, ioaseito\}@ihu.gr};\ %
\textsuperscript{3}Aristotle University of Thessaloniki, Greece,
{\footnotesize\ttfamily katsaros@csd.auth.gr};\ %
\textsuperscript{4}University of Thessaly, Greece,
{\footnotesize\ttfamily alekidis@uth.gr}}%
}}

\begin{document}
\maketitle
{\renewcommand{\thefootnote}{}\footnotetext{\copyright~2026 IEEE. Personal use of this material is permitted. Permission from IEEE must be obtained for all other uses, in any current or future media, including reprinting/republishing this material for advertising or promotional purposes, creating new collective works, for resale or redistribution to servers or lists, or reuse of any copyrighted component of this work in other works. This work has been accepted for publication at the 2026 IEEE International Conference on Cyber Security and Resilience (IEEE CSR 2026).}}

\begin{abstract}
Cyber resiliency in edge-IoT deployments is fundamentally an
economic problem: detection must keep critical processes operating
under attack, but defender resources (compute, bandwidth, operator
attention) are bounded.  Centralised cloud monitoring offers
expressive cross-device detection at prohibitive bandwidth cost;
purely edge-local monitoring is cheap but blind to coordinated
multi-device attacks where the asymmetric balance favours the
attacker.  We propose a lightweight hierarchical security-monitoring
framework, built on formal runtime-verification methods, that
occupies the practical middle ground at quantified cost.  Each edge device runs a lightweight TeSSLa stream
specification (size, payload validity, rate, and timestamp-drift
predicates) that emits a four-valued verdict per aggregation
window at sub-microsecond per-event cost; the gateway runs a
parametric first-order MonPoly monitor over the per-device verdict
streams at microsecond-scale per-verdict cost.  The edge-to-gateway
uplink carries roughly one Boolean per aggregation window per node,
orders of magnitude smaller than the raw packet stream.  The gateway
tier detects coordinated attack patterns that no single-node monitor
can see, shifting the asymmetric cost balance toward the defender.
Every alert carries a witness set naming the device, the monitor
tier, and the predicate that fired, providing an auditable record
of the decision.
We evaluate the framework on a container-host testbed spanning
nominal and attacker nodes across four attack classes (buffer
overflow, time spoofing, denial-of-service, and mixed
advanced-persistent-threat patterns), and describe the edge- and
gateway-tier specifications together with the cost-versus-coverage
trade-off as monitor levels are added.
\end{abstract}


\section{Introduction}
\label{sec:intro}

Edge-IoT deployments increasingly host security-relevant
decision-making at low-power devices: per-sensor anomaly detection,
local rate limiting, real-time intrusion classification, and
embedded perception on connected-vehicle platforms~\cite{chintri2026v2x}.
Two
architectural extremes are commonly used to enforce safety and
security properties on such deployments, and each has known failure
modes.

\emph{Edge-only monitors} are inexpensive and decide at line rate, but
are structurally blind to attacks that coordinate across devices: an
attacker that behaves nominally most of the time and bursts
intermittently is, by construction, invisible to any monitor that sees
only one node.  \emph{Cloud-only monitors} obtain the cross-node view
but pay for it in bandwidth (every packet shipped upward) and
detection latency, and they cannot enforce safety properties locally
at line rate.

This paper proposes a hierarchical security-monitoring framework
built on formal runtime-verification
methods~\cite{leucker2009brief}:
lightweight, single-property monitors at the edge and a single
parametric first-order monitor at the gateway, with a formal
composition operator between the two levels.  The framework is
instantiated over two mature open-source RV engines:
TeSSLa~\cite{tessla2018,tessla2024} for stream-based per-node monitoring with
timed asynchronous semantics, and MonPoly~\cite{monpoly2014} for
metric first-order temporal logic (MFOTL) at the gateway with
quantification over node identifiers.  The two levels emit four-valued
verdicts in the Bauer--Leucker--Schallhart
sense~\cite{blsemantics2011} and compose through a typed algebra that
records witnesses \emph{by level and by node}.  The architectural
result is that fast, deterministic decisions stay at the edge, the
expensive cross-node correlation runs only at the gateway, and each
blocked or flagged event carries an audit-oriented record of which level
and which node triggered the verdict.

We evaluate the framework on a 12-actor dataset comprising five
nominal nodes and seven attackers across four attack classes
(buffer overflow, time spoofing, denial-of-service, and mixed
\emph{Advanced Persistent Threat} (APT)).  An Advanced Persistent
Threat is a sustained, stealthy attack pattern in which one or
more actors maintain access to a system over an extended period
and coordinate intermittent violations across components so that
no single component, viewed in isolation, exhibits behaviour
unambiguously distinguishable from a noisy benign actor; the
attack signature emerges only from cross-component correlation
over a long enough window.  Three of the four attack classes
(overflow, time spoofing, denial-of-service) are caught by the edge
tier; the fourth (mixed APT, where three coordinating attackers each
behave nominally roughly 60\% of the time) is only presumptively
flagged per node at the edge ($\bot_p$) and confirmed as a
coordinated attack when the gateway tier's parametric correlation
across nodes is added.  This dataset motivates the hierarchical
framing: an attack class whose per-node evidence is individually
ambiguous and whose signature is confirmable only by cross-node
correlation.

\emph{Why this is hard for current single-level approaches.}
Edge-only stream-based RV (RTLola~\cite{rtlola2019},
TeSSLa~\cite{tessla2018}) has rich temporal semantics but operates on
a single trace and cannot express
\emph{parametric correlation across $N$ traces from independent
devices}.  At the gateway, MonPoly's MFOTL~\cite{monpoly2014} offers
exactly the quantification we need (``$\forall n, m, k$.\ three nodes
exhibiting intermittent violations within a 60\,s window''), but
running MonPoly on every raw packet from $N$ devices is bandwidth- and
compute-prohibitive.  The contribution of this paper is a clean split:
TeSSLa on the edge against raw packet streams, MonPoly at the gateway
against the (small) verdict streams emitted by each edge monitor, and
a formal algebra to compose the resulting verdicts.

The two-level split builds on hierarchical CPS-RV~\cite{strel2017,sastl2020};
the technical contribution is its instantiation with parametric
first-order MFOTL at the gateway tier, which catches the mixed-APT
patterns existing edge-IoT monitors miss.

\emph{Cost asymmetry.}
The framework forces the attacker to pay coordination cost across
$\geq\!3$ devices while the defender pays microsecond-scale cost per
event at the edge and per verdict at the gateway, with witness sets
at $O(1)$ per alert as the audit by-product.  The uplink is roughly
one Boolean per node per aggregation window, orders of magnitude
smaller than the raw stream a cloud correlator would require.

\emph{Threat model.}
We assume an adversary capable of compromising up to $K$ of $N$
edge devices and controlling their outbound packet streams within
physical-channel limits.  Real-world precedents such as the 2015
Ukrainian power-grid attack~\cite{ukraine2015} show that coordinated
cross-device APT is a realistic capability against edge-IoT-like fleets.  The
attacker may keep per-device behaviour intermittent to evade
single-node thresholds and may synchronise across the compromised
set.  \emph{Monitor integrity.}  The L1 monitor is an isolated sidecar
(separate process/namespace) beside, not inside, the network-facing
application, its integrity assumed under platform attestation.
Subverting an L1 monitor only disables local detection on that device,
the single-node-blind case the gateway already covers, since one
compromised edge cannot suppress the cross-device $\bot_p$ pattern.

Out of scope (assumed handled by attestation/isolation):
supply-chain compromise of the L1/L2 binaries, tampering with the
gateway-observed timestamp, and direct interference with the sidecar
L1 process.

\emph{Contributions.}
We first characterise \emph{detectability}: which attack classes are
decidable locally at the edge (overflow, time-spoofing, DoS) and which
require cross-device correlation at the gateway (coordinated
mixed-APT).  This local/global split is what the two-tier design
exploits.  Concretely, we give a two-level engine-agnostic
specification with a four-valued composition algebra and witness sets
(Section~\ref{sec:spec}), a concrete TeSSLa + MonPoly realisation
(Section~\ref{sec:arch}), and a measured evaluation on a 12-actor
testbed (extended in a 15-actor second-generation replay) in
Section~\ref{sec:eval}.

\emph{Motivating example: which class needs the gateway.}
The controlled 12-actor dataset (five nominal nodes, seven attackers,
four attack classes on a Docker testbed; schema and per-actor
configuration in Section~\ref{sec:eval:data}) isolates which class
requires the gateway tier.  The single-node classes (overflow, time
spoofing, DoS) each violate one invariant on one device and are caught
by a single-property monitor there.  The three mixed-APT actors
(\textsc{node-8/9/10}) instead behave nominally $\approx\!60\%$ of the
time and burst independently the rest; individually they emit
predominantly $\bot_p$ (one $\bot$).  The signal is the
\emph{cross-node correlation} (three independent $\bot_p$ verdicts
within a 60\,s window), which only a parametric quantifier
(FORALL $n$, EXISTS $m,k$, bounded-past ONCE) can express.

The empirical L1 verdict distribution grounding this observation
appears in Table~\ref{tab:walkthrough}.

\section{Specification and Composition}
\label{sec:spec}

A framework specification is a pair $\Phi = \langle \varphi_{L1},
\varphi_{L2} \rangle$, given in past-time first-order temporal
logic so the properties are engine-independent.  We realise them
in Section~\ref{sec:arch} as a TeSSLa~\cite{tessla2018}
program (L1, Listing~\ref{lst:tessla}) and a
MonPoly~\cite{monpoly2014} MFOTL program (L2,
Listing~\ref{lst:monpoly}); RTLola~\cite{rtlola2019} at L1 and
DejaVu~\cite{havelund2017dejavu} at L2 are alternative backends.
The framework consumes two streams: typed packet events
$e_i = \langle \mathsf{turn}_i, \mathsf{actor}_i, \mathsf{tool}_i,
\mathsf{args}_i \rangle$ at the edge (JSON schema in
Section~\ref{sec:eval:data}) and per-window verdict events
$v_j = \langle \mathsf{actor}_j, \mathsf{verdict}_j, \mathsf{turn}_j
\rangle$ at the gateway, with $\mathsf{verdict} \in V_4 =
\{\top, \top_p, \bot_p, \bot\}$.

\subsection{Two-level specification}
\label{sec:spec:two-level}

\paragraph{Parameters}  Throughout, $\sigma_{\max}$ is the
declared per-event buffer bound (30~bytes in our testbed),
$\rho_{\min}$ the minimum acceptable inter-arrival
(0.5~s), $\tau_{\mathit{jit}}$ the per-actor turn-monotonicity
tolerance (1.0~s), $W_{\mathit{agg}}$ the L1 aggregation window
(5~s), $N_{\mathit{hard}}$ the per-window violation count above
which L1 emits a definite block (3), and $W_{\mathit{past}}$ the
L2 bounded past-time window (60~s).  These are deployment-specific
operating points calibrated to the testbed's nominal $\sim$2\,s
inter-arrival, not universal constants: on a different fleet they must
be recalibrated against the target's nominal traffic, and the
sensitivity of detection and FPR to $\rho_{\min}$ and $W_{\mathit{agg}}$
is reported in Section~\ref{sec:eval}.

\paragraph{Local specification $\varphi_{L1}$ (per-device, edge)}
For each device $d$, let $\mathsf{events}_d$ denote the per-actor
event stream and let $\mathsf{prev\_turn}(e)$ denote the
\textsf{turn} of the most recent event from the same actor before
$e$.  Four single-property predicates run per event:
{\small
\begin{align*}
\mathsf{size\_ok}(e)  &\equiv\; e.\mathsf{actual\_sent}
                              \leq e.\mathsf{buffer\_limit} \\
\mathsf{fuzz\_ok}(e)  &\equiv\; e.\mathsf{actual\_sent} > 0 \\
\mathsf{mono\_ok}(e)  &\equiv\; e.\mathsf{turn} \geq
                              \mathsf{prev\_turn}(e) - \tau_{\mathit{jit}} \\
\mathsf{rate\_ok}(e)  &\equiv\; e.\mathsf{turn}
                              - \mathsf{prev\_turn}(e) \geq \rho_{\min} \\
\mathsf{event\_ok}(e) &\equiv\; \mathsf{size\_ok}(e) \wedge \mathsf{fuzz\_ok}(e) \\
                      &\phantom{{}\equiv{}}\; \wedge\, \mathsf{mono\_ok}(e) \wedge \mathsf{rate\_ok}(e).
\end{align*}}
A violation of each predicate maps to one single-node attack
class: $\mathsf{size\_ok}$/overflow, $\mathsf{fuzz\_ok}$/fuzzing,
$\mathsf{mono\_ok}$/time-spoofing, $\mathsf{rate\_ok}$/DoS.

\paragraph{L1 four-valued window verdict}
For each aggregation window $w$ of length $W_{\mathit{agg}}$
and each device $d$, let
\(
\mathit{nv}(d,w) = \big|\{e \in \mathsf{events}_d \cap w :
                       \neg\,\mathsf{event\_ok}(e)\}\big|
\)
be the number of property-violating events, and
\(
\mathit{sh}(d,w) = \big|\{e \in \mathsf{events}_d \cap w :
                          \mathit{iat}(e) < 1.5\rho_{\min}\}\big|
\)
the number of events whose inter-arrival is within 50\,\% of the
rate threshold.  Then
\begin{equation*}
\varphi_{L1}(d, w) \;\;=\;\;
\begin{cases}
\bot       & \text{if } \mathit{nv}(d,w) \geq N_{\mathit{hard}} \\
\bot_p     & \text{if } 1 \leq \mathit{nv}(d,w) < N_{\mathit{hard}} \\
\top_p     & \text{if } \mathit{nv}(d,w) = 0 \wedge
                         \mathit{sh}(d,w) \geq N_{\mathit{hard}} \\
\top       & \text{otherwise.}
\end{cases}
\end{equation*}
In plain terms: $\bot$ when $\geq\!N_{\mathit{hard}}$ events
violated any predicate, $\bot_p$ for one or two violations,
$\top_p$ for a near-rate-limit burst with no violations, and
$\top$ otherwise.

\paragraph{Global specification $\varphi_{L2}$ (gateway)}
The L2 monitor consumes the per-actor verdict events
$\mathsf{punsat}(n)$ emitted whenever $\varphi_{L1}(n,w) = \bot_p$
for some window $w$, and detects coordinated-APT correlation across
$N$ devices.  The baseline rule is
\begin{equation*}
\begin{aligned}
\forall n\,\exists m,k.\;\; & (n,m,k \text{ distinct}) \\
\wedge\;\; & \textstyle\bigwedge_{x\in\{n,m,k\}} \Diamond^{-}_{[0,W_{p}]}\mathsf{punsat}(x) \\
\to\;\; & \mathsf{coordinated\_apt\_alert}(n,m,k),
\end{aligned}
\end{equation*}
where $W_{p} \equiv W_{\mathit{past}} = 60$\,s and
$\Diamond^{-}_{[a,b]}$ is the bounded-past-time operator.  The
rule fires when three distinct devices each emitted $\geq\!1$
$\bot_p$ verdict in the past $W_{p}$, the cross-device pattern
no single-device monitor can see.

The \emph{robust variant} $\varphi_{L2}^{(k)}$ replaces
$\mathsf{punsat}$ with $\mathsf{punsat}_k$, requiring each
contributing actor to have emitted $\geq k$ presumptive-unsat
verdicts in the past $W_{\mathit{past}}$ (see
Listing~\ref{lst:monpoly}); this suppresses alerts driven by
sparse isolated noise.  We adopt $k{=}3$ on the 12-actor trace
and $k{=}2$ on the 15-actor replay
(Section~\ref{sec:eval:secondgen}).

\subsection{Composition algebra}
\label{sec:spec:compose}

Each level emits $v_\ell \in V_4$.  The composed verdict
$v = v_{L1} \otimes v_{L2}$ is the pointwise meet on the
Bauer--Leucker--Schallhart lattice ($\bot$ dominates, $\top$
confirms; e.g.\ $\bot_p \otimes \bot = \bot$,
$\bot_p \otimes \bot_p = \bot_p$).  Each verdict carries a
witness set $W$ tagged by level and node; composition unions
witnesses.  The split is asymmetric: single-node attacks
(\textsc{node}-5/6/7) are caught at the edge
($v_{L1}\in\{\bot_p,\bot\}$); coordinated APT (\textsc{node}-8/9/10)
yields predominantly $v_{L1}{=}\bot_p$ per device and relies on
$v_{L2}{=}\bot$ to fire.

\noindent\textbf{Lemma 1 (composition soundness).}\;
\emph{If $\varphi_{L1}$ and $\varphi_{L2}$ are sound with respect to
their per-trace semantics, then $v_{L1} \otimes v_{L2} = \bot$
implies the joint trace violates $\varphi_{L1} \wedge \varphi_{L2}$,
and the witness set $W = W_{L1} \cup W_{L2}$ identifies a
violating sub-trace.}
\emph{Proof sketch.} By definition of the meet, $v_{L1} \otimes
v_{L2} = \bot$ requires $v_{L1} = \bot$ or $v_{L2} = \bot$.
Per-level soundness then provides a violating sub-trace tagged in
the corresponding witness set; the set-union closes the composition
case.  Associativity and commutativity of the lattice meet extend
the lemma to compositions of any finite arity.\hfill$\Box$

\section{Runtime Architecture}
\label{sec:arch}

This section gives one concrete realisation of the engine-agnostic
specification of Section~\ref{sec:spec:two-level}.  $\varphi_{L1}$
is realised as a TeSSLa stream program (Listing~\ref{lst:tessla})
and $\varphi_{L2}$ as a MonPoly MFOTL program
(Listing~\ref{lst:monpoly}); the parameters
($\sigma_{\max}, \rho_{\min}, \tau_{\mathit{jit}},
W_{\mathit{agg}}, N_{\mathit{hard}}, W_{\mathit{past}}$) are
those defined in Section~\ref{sec:spec:two-level}.

\paragraph{Deployment topology}
The L1 TeSSLa monitor runs in-process on each IoT device,
enforcing single-property invariants at line rate; the L2 MonPoly
monitor runs once at a local gateway one network hop above the
devices, consuming the per-actor verdict streams from every L1.
The split keeps per-packet workload on the device and gives the
gateway the cross-device view needed for parametric APT
detection over a one-Boolean-per-window-per-node uplink.  In the
evaluation (Section~\ref{sec:eval}) both tiers are realised as
separate Docker containers on a single host; per-event latency on
physical Pi hardware is left to a follow-up.

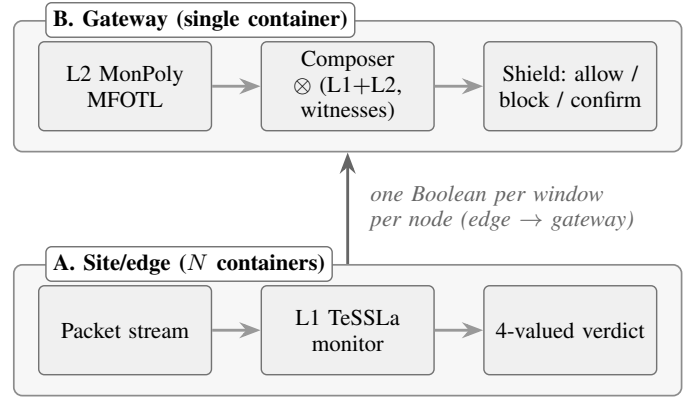
\begin{figure}[t]
\centering
\resizebox{\columnwidth}{!}{%
\begin{tikzpicture}[
  font=\footnotesize,
  cell/.style={draw=black!55, line width=0.5pt, rounded corners=2pt,
               fill=black!6, inner sep=2pt, text width=20mm,
               minimum height=11mm, align=center},
  band/.style={draw=black!45, line width=0.6pt, rounded corners=3pt,
               fill=black!3, inner ysep=2.5mm, inner xsep=3mm},
  tab/.style={draw=black!55, line width=0.5pt, rounded corners=2pt,
              fill=white, font=\footnotesize\bfseries, inner sep=2.5pt},
  flow/.style={-Stealth, black!60, line width=1pt}
]
\node[cell] (gl2) {L2 MonPoly MFOTL};
\node[cell, right=6mm of gl2]   (gcomp)   {Composer $\otimes$ (L1$+$L2, witnesses)};
\node[cell, right=6mm of gcomp] (gshield) {Shield: allow / block / confirm};
\draw[flow, black!40] (gl2)   -- (gcomp);
\draw[flow, black!40] (gcomp) -- (gshield);
\begin{scope}[on background layer]
  \node[band, fit=(gl2)(gcomp)(gshield)] (gw) {};
\end{scope}
\node[tab, anchor=west] at ([xshift=4mm]gw.north west)
   {B.\ Gateway (single container)};
\node[cell, below=19mm of gl2]     (epkt) {Packet stream};
\node[cell, below=19mm of gcomp]   (el1)  {L1 TeSSLa monitor};
\node[cell, below=19mm of gshield] (ever) {4-valued verdict};
\draw[flow, black!40] (epkt) -- (el1);
\draw[flow, black!40] (el1)  -- (ever);
\begin{scope}[on background layer]
  \node[band, fit=(epkt)(el1)(ever)] (edge) {};
\end{scope}
\node[tab, anchor=west] at ([xshift=4mm]edge.north west)
   {A.\ Site/edge ($N$ containers)};
\draw[flow] (edge.north) -- (gw.south)
   node[midway, right=1.5mm, align=left, font=\footnotesize\itshape]
   {one Boolean per window\\per node (edge $\to$ gateway)};
\end{tikzpicture}%
}
\caption{Runtime architecture. Each of $N$ edge containers runs an
identical L1 TeSSLa pipeline that turns its raw packet stream into a
four-valued verdict per window; a single gateway container hosts the
L2 MonPoly MFOTL monitor, the composer (pointwise verdict fusion with
per-level witness union), and the shield. The uplink carries one
Boolean per window per node, the bandwidth-saving property that makes
the architecture deployable.}
\label{fig:arch}
\end{figure}

The runtime follows a sidecar pattern: an L1 process runs on each
edge device alongside the network stack and inspects the device's
outbound packet stream; a single L2 process runs at the gateway over
the aggregated per-node verdict streams.  Figure~\ref{fig:arch} shows
the flow.

\subsection{Event canonicalisation}
\label{sec:arch:canon}

The testbed-generated event log is JSON-encoded (one event per
line).  A small Python adapter performs three jobs:
(i) it parses the JSONL stream into typed packet events; (ii) it
groups events by \textsf{actor}, producing a per-actor input stream
to the L1 monitor; and (iii) it translates the L1 monitor's
four-valued verdicts into a MonPoly log of single-argument events,
one per (actor, window) pair.  The MonPoly signature declares one
actor-keyed predicate per L1 verdict value (\textsf{sat},
\textsf{psat}, \textsf{punsat}, \textsf{unsat}), the gateway-side
\textsf{edge\_block}, and the ternary alert predicate
\textsf{coordinated\_apt\_alert} with its robust variant.

\subsection{Concrete L1 and L2 specifications}
\label{sec:arch:l1}

\emph{L1 (edge).}
Listing~\ref{lst:tessla} reproduces the per-node TeSSLa specification.
Four single-property invariants run in parallel; the per-event
conjunction is aggregated over a five-second window into a four-valued
verdict that is sampled and shipped to the gateway.

\begin{lstlisting}[float=tbp,language=tessla,caption={L1 TeSSLa specification
(per-actor, edge); parameters as in Section~\ref{sec:spec:two-level}.},
label=lst:tessla]
in evt : Events[Event]     // per-actor; size/fuzz/mono/rate_ok as in Sec. II
def iat = lift2((p,c) => float(c-p), last(evt.turn,evt), evt.turn)
def event_ok = lift4((a,b,c,d) => a&&b&&c&&d, size_ok,fuzz_ok,mono_ok,rate_ok)
def nv = slidingCount(lift(b => !b, event_ok), AGG_WIN_S)   // violations/window
def sh = slidingCount(lift(d => d < 1.5*MIN_IAT_S, iat), AGG_WIN_S)
def local_verdict = lift2((nv,sh) =>          // 3=UNSAT 2=PUNSAT
   if nv>=N_HARD then 3 else if nv>=1 then 2   // 1=PSAT  0=SAT
   else if sh>=N_HARD then 1 else 0, nv, sh)
send (actor, local_verdict) to gateway
\end{lstlisting}

\emph{L2 (gateway).}
Listing~\ref{lst:monpoly} reproduces the gateway-tier MFOTL
specification.  \textsf{edge\_block\_rule} echoes a definite L1 block
to the composer; \textsf{coordinated\_apt\_robust\_rule} is the
coordinated-APT alert, requiring each contributing node to emit
$\bot_p$ at least $k$ times within the 60-second window via the
auxiliary predicate \textsf{punsat\_twice} (shown for $k{=}2$; the
recommended operating point on the 12-actor trace is $k{=}3$,
Section~\ref{sec:eval}).  This suppresses the sparse single-event
clean-node false-positives reported in Section~\ref{sec:eval} on
cross-node alerts.  (The non-robust baseline drops
\textsf{punsat\_twice} for a plain \textsf{punsat}, $k{=}1$.)

\begin{lstlisting}[float=tbp,language=mfotl,caption={L2 MonPoly MFOTL
specification (gateway). \textsf{ONCE[0,60s]} is the bounded-past
operator; quantifiers range over observed nodes.},
label=lst:monpoly]
edge_block_rule :=
   FORALL n. unsat(n) IMPLIES edge_block(n)
punsat_twice(n) := punsat(n) AND (ONCE[1s,60s] punsat(n))
coordinated_apt_robust_rule :=
   FORALL n, m, k.
     (n != m AND n != k AND m != k
       AND (ONCE[0,60s] punsat_twice(n))
       AND (ONCE[0,60s] punsat_twice(m))
       AND (ONCE[0,60s] punsat_twice(k)))
     IMPLIES coordinated_apt_alert_robust(n, m, k)
\end{lstlisting}

\paragraph*{Composer and shield}
The composer applies the lattice pointwise on each L1/L2 verdict
pair, unions the per-level witness sets, and emits
$\langle\mathsf{verdict}, W, \mathsf{evidence}\rangle$.  The shield
maps $\bot$ to block, $\bot_p$ to require-confirm, $\top_p$ to
allow-with-logging, and $\top$ to allow; every event-verdict pair
can be appended to a hash-chained audit log (mechanism described,
not evaluated).

\section{Evaluation and Methodology}
\label{sec:eval}

The methodology is described in enough detail for independent
reproduction on the 12-actor dataset.

\subsection{Dataset and testbed}
\label{sec:eval:data}

The controlled 12-actor dataset introduced in
Section~\ref{sec:intro} is the evaluation surface.  All traces are generated on a Docker-based
testbed in which each node is a container emulating a
Raspberry-Pi-class edge device, with all containers communicating
with a separate gateway container over a virtual network;
ground-truth labels are recorded at the generator.  The containers
run on a single workstation-class host (Apple M4 Pro, 12 cores,
8 performance plus 4 efficiency, 24\,GB RAM, macOS~15.3)
without per-container CPU or memory caps: the testbed emulates the
Raspberry-Pi-4 device class (quad-core Cortex-A72, 1--8\,GB RAM) at
the software and protocol level rather than in its resource envelope.
The per-event L1 costs reported below are therefore host-relative
upper-performance figures; a physical-Raspberry-Pi timing study,
where the monitors share the device with the primary edge workload,
is a follow-up (Section~\ref{sec:discussion}).

\paragraph{Event schema}  Each node emits one JSON
\textsf{tool\_call} event per line to a shared gateway log, with
fields \textsf{turn} (gateway-observed timestamp in seconds),
\textsf{actor} (originating node), \textsf{tool}
(\textsf{safe\_send}, \textsf{vulnerable\_send}, or
\textsf{time\_spoof\_send}), and \textsf{args} (\textsf{actual\_sent}
payload size in bytes and \textsf{buffer\_limit}, the declared bound,
30 bytes throughout).  The three single-property violations (oversize
\textsf{actual\_sent}, non-monotone \textsf{turn}, short
inter-arrival) are directly computable from these fields; the adapter
(Section~\ref{sec:arch:canon}) groups events by \textsf{actor} and
emits per-actor packet streams to L1 and verdict events to L2.

\paragraph{Per-actor configuration}  The twelve actors are configured
as follows.
\textsc{node-1}--\textsc{node-4} and \textsc{raspi-node-1}
(\emph{clean}): nominal baseline, mean inter-arrival $\approx 2$\,s,
payload uniform in $[10,30]$ bytes, always \textsf{safe\_send}.
\textsc{node-5} and \textsc{raspi-node-2} (\emph{buffer overflow}):
baseline timing, payload uniform in $[35,60]$ with
\textsf{actual\_sent} $>$ \textsf{buffer\_limit} on every overflow
event (\textsf{vulnerable\_send}).
\textsc{node-6} (\emph{time spoofing}): nominal size and rate, but the
\textsf{turn} field is back-jumped on a fraction of events, producing
non-monotone per-actor timestamps.
\textsc{node-7} (\emph{denial-of-service}): legal sizes, sustained
burst at $\approx 0.2$\,s inter-arrival (an order of magnitude faster
than baseline).
\textsc{node-8}--\textsc{node-10} (\emph{mixed APT}): nominally clean
$\approx$60\% of the time, the remaining 40\% a mixture of
\textsf{vulnerable\_send} and \textsf{time\_spoof\_send} sampled
independently per node.

The per-class L1 verdict breakdown is read directly from
Table~\ref{tab:walkthrough}: single-node classes (overflow,
ts-spoof, DoS) cross the L1 thresholds and yield $\bot_p$/$\bot$,
while the three mixed-APT nodes contribute only eleven sparse $\bot_p$
(and one $\bot$) that no node resolves alone but that jointly trigger
the L2 rule.

Three trace bundles, each in three runs for within-class variance:
$D_{\textsc{single}}$ (one trace per single-node class plus clean
actors, validates L1), $D_{\textsc{apt}}$ (nodes 8--10 mixed-APT plus
clean nodes 1--4, validates L2), and $D_{\textsc{benign}}$ (nominal
traces from all twelve actors, measures FPR).

\subsection{Baselines}
\label{sec:eval:baselines}

We compare the framework against five baselines drawn from the
edge-network-monitoring tradition, all on the same 5-second
aggregation windows as L1 so detection numbers are directly
comparable.
\emph{No monitor}: passthrough, giving the upper bound on ASR and the
lower bound on overhead.
\emph{B1 token-bucket}: a per-actor bucket of capacity 3 refilling at
$1$ token$/\rho_{\min}$, window flagged iff any event arrives with an
empty bucket (rate signal only).
\emph{B2 size cap}: $\mathsf{actual\_sent}\leq\mathsf{buffer\_limit}$
per event, flagged iff any event violates (size signal only).
\emph{B3 naive AND}: all three L1 predicates evaluated per event
without four-valued aggregation, flagged iff any event fails
(isolates the aggregation step).
\emph{L1 only}: edge-only monitoring with four-valued aggregation
(Section~\ref{sec:spec}).
Our \emph{L1$+$L2 (composed)} monitor (Section~\ref{sec:arch}) uses
the L2 robust variant ($k=3$) as the recommended operating point.

\subsection{Metrics and protocol}

We report per-attack-class detection rate (fraction of ground-truth
attack windows emitting a non-\textsf{sat} verdict), clean-actor
false-positive rate, and per-event L1 / per-verdict L2 cost.  Each
replay is offline: L1 runs per actor, its verdicts feed the L2 rule,
and the composed verdicts (with witness sets) are compared to
ground-truth labels.

\subsection{Measured results}

\paragraph{Per-actor L1 verdict distribution}
Table~\ref{tab:walkthrough} reports L1 verdicts per actor
(359 events, 76 five-second windows; $\rho_{\min}{=}0.5$\,s,
$\tau_{\mathit{jit}}{=}1$\,s, $N_{\mathit{hard}}{=}3$).

\begin{table}[t]
\centering
\caption{Per-actor L1 verdict counts (5\,s windows); attack-node
columns bold (N$=$node, R$=$raspi-node). Mixed-APT actors N8--N10
emit the sparse $\bot_p$ that L2 correlates.}
\label{tab:walkthrough}
\renewcommand{\arraystretch}{1.0}
\setlength{\tabcolsep}{3pt}
\footnotesize
\begin{tabular}{@{}l *{12}{c}@{}}
\toprule
 & N1 & N2 & N3 & N4 & \textbf{N5} & \textbf{N6} & \textbf{N7}
   & \textbf{N8} & \textbf{N9} & \textbf{N10} & R1 & \textbf{R2} \\
\midrule
total     & 7 & 7 & 6 & 6 & 6 & 4 & 7 & 6 & 7 & 3 & 8 & 9 \\
$\top$    & 6 & 6 & 4 & 4 & 0 & 2 & 0 & 2 & 2 & 0 & 6 & 0 \\
$\top_p$  & 0 & 0 & 0 & 0 & 0 & 0 & 0 & 0 & 0 & 0 & 0 & 0 \\
$\bot_p$  & 1 & 1 & 2 & 2 & 5 & 2 & 0 & 3 & 5 & 3 & 2 & 1 \\
$\bot$    & 0 & 0 & 0 & 0 & 1 & 0 & 7 & 1 & 0 & 0 & 0 & 8 \\
\bottomrule
\end{tabular}
\end{table}

Single-node attacks cross L1 thresholds reliably; mixed-APT actors
only emit sparse $\bot_p$ and require L2 cross-stream correlation.
Clean-actor FPs (8/34, $\approx\!24\%$) trace to sub-second
timestamp granularity (Section~\ref{sec:discussion}); the L2
robust variant suppresses their impact on cross-node alerts.

\paragraph{Aggregate detection, FPR, and overhead}
Table~\ref{tab:results} summarises detection, FPR, and per-event
cost.  Ground truth: five clean actors (34 windows) and seven
attackers across four classes (overflow, ts-spoof, DoS, APT;
42 windows).

\begin{table}[t]
\centering
\caption{Detection (attack windows flagged$\uparrow$) and
FPR$\downarrow$ per configuration, on the 12-actor trace and the
15-actor replay. A dash marks an attack class absent from that trace.}
\label{tab:results}
\renewcommand{\arraystretch}{1.0}
\setlength{\tabcolsep}{3pt}
\footnotesize
\begin{tabular}{@{}l c c c c c c c c@{}}
\toprule
 & \multicolumn{7}{c}{Detection (per attack class)} & FPR \\
\cmidrule(lr){2-8}
Configuration & ovfl & tsp & DoS & APT & stlth & fuzz & puls & \\
\midrule
\multicolumn{9}{@{}l}{\emph{12-actor trace} (359 events, 76 windows)}\\
No monitor               & 0/15  & 0/4 & 0/7 & 0/16  & -- & -- & -- & 0/34 \\
B1 token-bucket          & 0/15  & 2/4 & 7/7 & 6/16  & -- & -- & -- & 0/34 \\
B2 size cap              & 15/15 & 0/4 & 0/7 & 8/16  & -- & -- & -- & 0/34 \\
B3 naive AND             & 15/15 & 2/4 & 7/7 & 12/16 & -- & -- & -- & 8/34 \\
L1 only                  & 15/15 & 2/4 & 7/7 & 12/16 & -- & -- & -- & 8/34 \\
\textbf{L1+L2 ($k{=}3$)} & \textbf{15/15} & \textbf{2/4} & \textbf{7/7}
                         & \textbf{14/16} & -- & -- & -- & \textbf{8/34} \\
\midrule
\multicolumn{9}{@{}l}{\emph{15-actor replay} (363 events, 65 windows;
+stealth, fuzz, pulsing DoS)}\\
No monitor               & 0/5 & 0/3 & 0/6 & 0/15  & 0/5 & 0/6 & 0/2 & 0/23 \\
L1 only                  & 5/5 & 2/3 & 6/6 & 12/15 & 2/5 & 5/6 & 2/2 & 1/23 \\
\textbf{L1+L2 ($k{=}2$)} & \textbf{5/5} & \textbf{2/3} & \textbf{6/6}
                         & \textbf{14/15} & \textbf{2/5} & \textbf{5/6}
                         & \textbf{2/2} & \textbf{1/23} \\
\bottomrule
\end{tabular}
\end{table}

The baselines each cover at most one signal: B1 (rate) catches
DoS (7/7) and partial ts-spoof (2/4); B2 (size) catches overflow
(15/15); B3 (naive AND) matches L1 per window but is two-valued
and cannot drive the gateway's four-valued rule.  Only L1 covers
all three single-node classes simultaneously, and only L1+L2
catches the coordinated APT (14/16 vs.\ 12/16 for L1-only).

\paragraph{Sensitivity}  FPR is invariant to
$\rho_{\min} \in \{0.3, 0.5, 0.7\}$\,s: all eight clean-node PUNSAT
windows stem from event pairs with coincident integer-second
\textsf{turn} ($iat{=}0$) and fail the rate predicate at any positive
threshold (a sub-second timestamp would remove them,
Section~\ref{sec:discussion}).  Sweeping $\mathsf{AGG\_WIN\_S} \in
\{2,5,10,30\}$\,s, ts-spoof detection stays at 50\% (backward jumps
concentrate in two of four windows); shrinking to 2\,s drops FPR from
24\% to 14\% without harming per-class detection, widening to
10--30\,s raises it to 32--50\% as each FP taints a longer window.

\paragraph{L2 alert stream and the choice of $k$}  The baseline
rule fires 18 alerts, each mixing genuine attackers with at
least one clean-actor sub-second-collision FP; tightening to
$k{=}2$ yields 10 (2 true / 8 mixed) and $k{=}3$ yields 6, all
six containing three real attackers with no clean contributors on
this trace.  We adopt $k{=}3$ as the operating point in
Table~\ref{tab:results}; the L1+L2 union covers 14/16
attacker windows by recovering two L1-misses through L2's
60\,s past window.

\paragraph{Cost profile}
L1 costs 0.52\,$\mu$s per event (p95 0.67\,$\mu$s) and L2
$\sim$1\,$\mu$s per verdict, three orders of magnitude under the
sub-millisecond budget; under linear scaling this projects to under
1\,\% of one core at $N{=}1000$ (L2 cost at scale is unmeasured).
The L1$\to$L2 uplink is 200\,bits/s
at $N{=}1000$, $\sim$4 orders of magnitude below a raw-packet
cloud correlator.  Witness sets are $O(1)$ per verdict and $O(k)$
per alert; audit storage for the whole evaluation is kilobyte-scale.

\subsection{Second-generation replay}
\label{sec:eval:secondgen}

A 15-actor re-run (363 events, 65 windows) in which the
$\mathsf{fuzz\_ok}(e) \equiv e.\mathsf{actual\_sent} > 0$ predicate
(vacuous on the original trace) becomes active extends coverage to stealth
overflow, fuzzing, and pulsing DoS; its detection and FPR figures
are the lower panel of Table~\ref{tab:results}.
All three new classes are detected by the existing L1 primitives
without architectural change (size predicate for stealth overflows,
$\mathsf{fuzz\_ok}$ for fuzzing, rate predicate for the pulsing
burst pattern); the false-positive rate drops to 1/23
($\approx\!4\%$) on this fresh trace because the same-second-collision
artifacts of the original log are absent, confirming the
data-collection-layer diagnosis (Section~\ref{sec:discussion}); and
the robust L2 variant reaches all-true alerts earlier, at $k{=}2$
(6 alerts, all true) rather than $k{=}3$ (3 alerts), against 15
baseline alerts (7 true / 8 mixed), so the recommended operating
point is data-dependent.

\section{Related Work}
\label{sec:related}

STREL~\cite{strel2017} and SaSTL~\cite{sastl2020} formalise
hierarchical spatio-temporal monitoring; we reuse the
decomposition pattern but replace propositional spatial operators
with parametric first-order MFOTL for cross-node correlation.
Stream-based embedded monitors, including the recent TeSSLa
ecosystem~\cite{tessla2024}, share the L1 design point: small,
deterministic, per-trace; advisor-monitored shields such as safe-RL
shielding~\cite{shielding2018} wrap an unsafe
policy in a monitor, related to the $\bot \to$ block /
$\bot_p \to$ require-confirm shield in Section~\ref{sec:arch}.
Temperekidis et al.~\cite{temperekidis2022fmirv} embed runtime
monitors in FMI-based co-simulation, a design-time analogue to
our deployment-time edge tier; at L2 we prefer
MonPoly~\cite{monpoly2014} over
DejaVu~\cite{havelund2017dejavu} for its native metric intervals.
Signature- and learning-based IoT IDS and uncertainty-aware
classifier monitoring~\cite{lukina2021into} are complementary to L1: a learned classifier
could populate $\bot_p$ without changing the L2 specification, and
the learned component itself can be formally certified via
neural-network equivalence checking~\cite{eleftheriadis2022nnequiv}.
Established resiliency and least-privilege
frameworks~\cite{nist800160v2} catalogue
defensive techniques but stop short of numerical cost models for
specific defender primitives; our framework instantiates one with
explicit costs against an attack class whose mounting-cost is
super-linear in compromised devices, giving inputs for
deployment-time cost-benefit analysis.

\section{Discussion}
\label{sec:discussion}

\paragraph{Limitations}  The 12-actor dataset is synthetic and the
four attack classes are a closed set; real-world mixes are
heavier-tailed and would need additional L1 predicates or threshold
tuning.  Replay over logged streams leaves detection-rate results
algorithmic, and the robust order $k$ is input-dependent ($k{=}3$ on
the 12-actor trace, $k{=}2$ on the cleaner replay).  The gateway is a
single point of evaluation; multi-gateway aggregation and concept
drift on the packet distribution are left to future work.  All eight
clean-actor false positives stem from event pairs with coincident
integer-second \textsf{turn} ($iat{=}0$), an artifact of the log's
one-second resolution that a sub-second timestamp would remove.
\emph{Adaptive evasion.}  Our adversary is non-adaptive; one knowing
$W_{\mathit{past}}$ and $k$ could space bursts across the past-window
boundary or thin them below $k$.  Randomised or overlapping windows,
adaptive $k$, and CUSUM-style accumulation are candidate mitigations,
and evaluating against adaptive, multi-stage adversaries is the main
open item.

\paragraph{Future directions}  The priorities are a physical
Raspberry-Pi timing study, a live closed-loop deployment (where the
same temporal-logic specifications that drive the monitors can also
synthesise the corrective controller~\cite{dang2023cegil}, closing a
monitor-to-control loop), and hardening against adaptive adversaries.  Longer term, multi-gateway aggregation, an optional
cloud tier for cross-gateway semantic review, and automated
property-catalogue synthesis from natural-language manuals would
extend coverage without re-engineering the monitors.

\section{Conclusion}
\label{sec:conclusion}

This paper presented a hierarchical security-monitoring framework
for edge-IoT, grounded in formal runtime-verification methods:
TeSSLa at the edge composed with MonPoly MFOTL at the gateway via
a typed four-valued algebra over witness sets.  The contribution
is the vertical composition, not the individual monitors.  On the 12-actor testbed (sub-microsecond per-event L1,
microsecond per-verdict L2, one-Boolean-per-window-per-node uplink
that is four orders of magnitude smaller than the raw packet
stream a cloud correlator would need) the composed monitor catches
every single-node attack window any baseline catches plus the
coordinated mixed-APT class no single-tier baseline can express.
A second-generation 15-actor replay (one extra L1 predicate; no
architectural change) extends coverage to stealth overflow,
fuzzing, and pulsing DoS, and the clean-actor FPR drops to
$\approx\!4\%$ on the fresh trace.  Its quantified cost figures make
it a deployable baseline against which heavier ML- or LLM-based
approaches to edge-IoT resilience can be evaluated.

\section*{Acknowledgment}
This work has received funding from the European Union's Digital
Europe Programme under grant agreement No~101190251 (CYBERGUARD).

\bibliographystyle{IEEEtran}
\bibliography{references}

\end{document}